\documentclass{INTERSPEECH2023}

\usepackage[pdfauthor={Tamas Gabor Csapo, Frigyes Viktor Arthur, Peter Nagy, Adam Boncz},
                pdftitle={Towards Ultrasound Tongue Image prediction from EEG during speech production},
                colorlinks=false,
               ]{hyperref}

\interspeechcameraready

 \usepackage[utf8]{inputenc}

\title{Towards Ultrasound Tongue Image prediction from EEG \\ during speech production}

\name{Tamás Gábor Csapó$^1$, Frigyes Viktor Arthur$^1$, Péter Nagy$^{2,3}$, Ádám Boncz$^3$}
\address{
  $^1$Department of Telecommunications and Media Informatics, \\
	Budapest University of Technology and Economics (BME), Budapest, Hungary \\
 $^2$Department of Measurement and Information Systems, BME, Budapest, Hungary \\
	$^3$Sound and speech perception Research Group,  Institute of Cognitive Neuroscience and Psychology, \\ 
	Research Centre for Natural Sciences, Budapest, Hungary}
\email{\{csapot,arthur\}@tmit.bme.hu, nagy.peter.ssprg@ttk.hu, adam.boncz@gmail.com}

\begin{document}

\maketitle
\begin{abstract}
Previous initial research has already been carried out to propose speech-based BCI using brain signals (e.g.~non-invasive EEG and invasive sEEG / ECoG), but there is a lack of combined methods that investigate non-invasive brain, articulation, and speech signals together and analyze the cognitive processes in the brain, the kinematics of the articulatory movement and the resulting speech signal.
In this paper, we describe our multimodal (electroencephalography, ultrasound tongue imaging, and speech) analysis and synthesis experiments, as a feasibility study. We extend the analysis of brain signals recorded during speech production with ultrasound-based articulation data. From the brain signal measured with EEG, we predict ultrasound images of the tongue with a fully connected deep neural network. The results show that there is a weak but noticeable relationship between EEG and ultrasound tongue images, i.e. the network can differentiate articulated speech and neutral tongue position.
\end{abstract}

\section{Introduction}

Brain-Computer Interfaces (BCIs) can allow computers to be controlled directly without physical activity. Augmentative and Alternative Communication (AAC) technologies (e.g.~BCI) can directly read brain signals to compensate for lost speech ability~\cite{Chang2020}. In the future, the use of speech neuroprostheses may help patients with neurological or speech disorders~\cite{Metzger2022}.
For recording the brain signal, several technologies are available: e.g., electroencephalography (EEG)~\cite{McFarland2017}, stereotactic deep electrodes (sEEG)~\cite{Verwoert2022}, intracranial electrocorticography (ECoG)~\cite{Buzsaki2012}, magnetoencephalography (MEG)~\cite{Dash2021}, Local Field Potential (LFP)~\cite{Buzsaki2012}. Among these brain signal recording methods, EEG may be the most suitable for BCI, as it is affordable, involves significantly less risk than invasive methods, and can be portable \cite{Casson2019}. Initial research has already been carried out to develop EEG and speech-based BCI~\cite{Krishna2020,Verwoert2022,Luo2022}, but this has not yet resulted in clearly intelligible speech. The reason is that EEG only measures the brain signal on the scalp; therefore, it is less accurate than invasive technologies. Using invasive methods, it has already been possible to create speech-like synthesized speech based on brain signals, e.g.~ECoG~\cite{Herff2015,Anumanchipalli2019} and sEEG~\cite{Angrick2021,Verwoert2022,Arthur2022a}, but due to the above disadvantage (primarily the invasive nature), the latter are not expected to be widespread.

\subsection{Brain signals and articulatory movement}

Articulatory movements have only sporadically been studied in parallel with brain signals during speech production, according to our knowledge. Articulation-based strategies require a two-step approach: a neural decoder for articulation and an articulation-to-speech model. There is a single study by Lesaja et al.~that investigates neural correlates of lip movements, and predicts lip-landmark position from ECoG input ~\cite{Lesaja2019}. Besides, the other related studies all use estimated articulatory data, i.e.~they take into account the articulatory information inferred from the speech signal or from textual contents~\cite{Guenther2009,Brumberg2011,Carey2017,Chartier2018,Anumanchipalli2019,Favero2022,LeGodais2022}. 

\begin{figure*}%[!b]
	%\centering
	 \begin{minipage}{0.7\linewidth}
			\centering
			% trims (crops) from left, bottom, right and top
		\includegraphics[trim=0.3cm 0.3cm 0.3cm 0.3cm, clip=true,width=\textwidth]{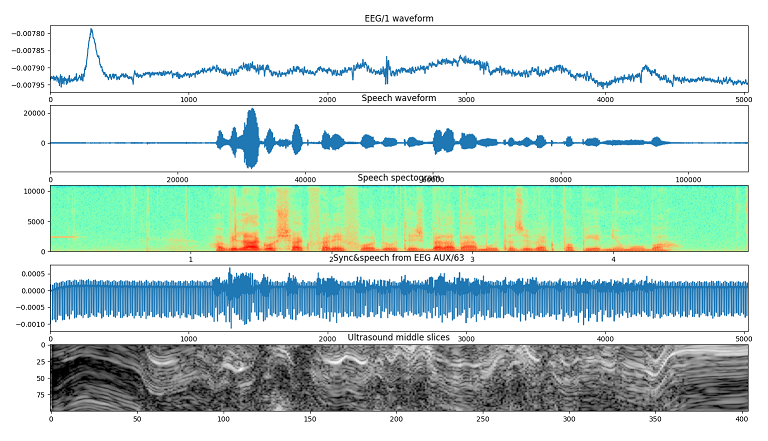}
			\tiny{Time (frame)}
			\caption{Example for synchronized EEG, speech, and ultrasound tongue imaging recordings. a) EEG / 1st channel, b) speech signal, c) speech spectrogram, d) ultrasound synchronization signal and speech signal (EEG on AUX), e) temporal change of the center line of UTIs.}
			\label{fig:eeg_sync}
		\end{minipage}
		\hspace{3mm}
	\begin{minipage}{0.27\linewidth}
		\centering
		\includegraphics[width=0.82\textwidth]{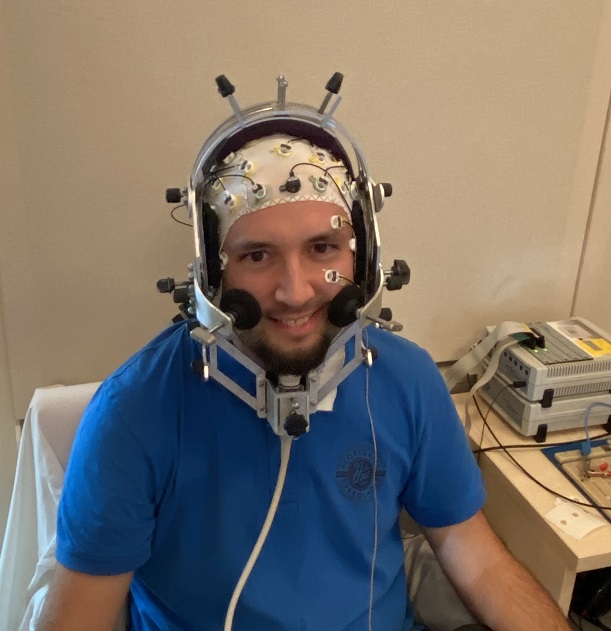}
	%\hspace{2mm}
		\includegraphics[width=0.82\textwidth]{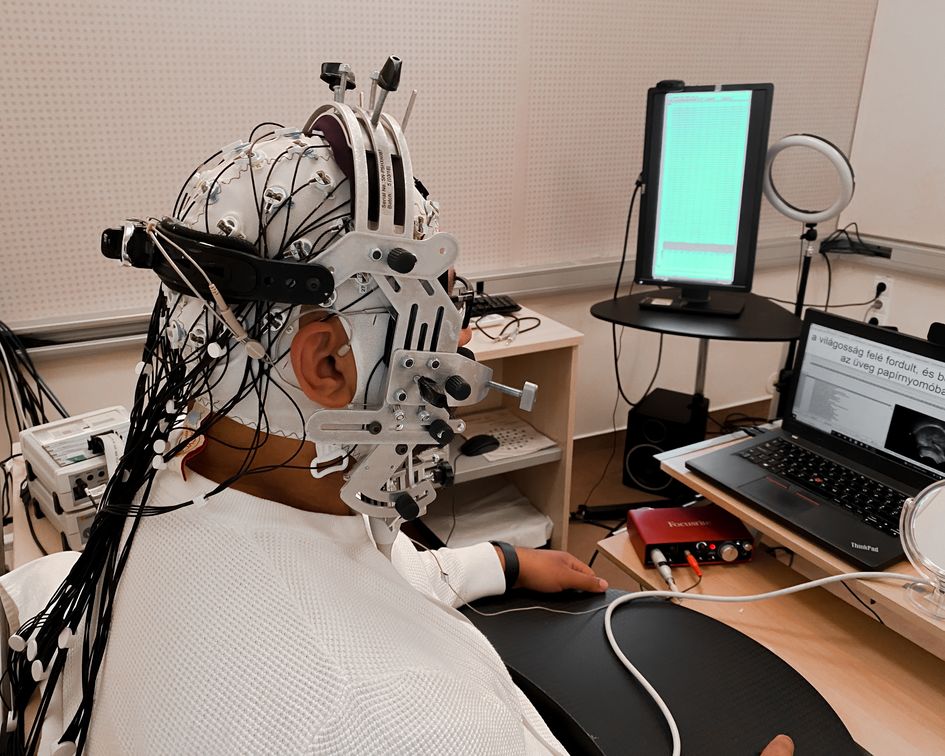}
		\caption{Recording setup: EEG, ultrasound tongue imaging with a headset, microphone and webcam.}
		\label{fig:felvetel}
	\end{minipage}
\end{figure*}

As early as a decade ago, Guenther et al.~suggested using articulatory information in speech-based BCIs~\cite{Guenther2009}. They state that articulatory data might be easier and more useful to predict from brain data than formant frequencies. In a follow-up study~\cite{Brumberg2011}, they classified American English phonemes from intracortical microelectrode recordings, considering the articulatory characteristics of vowels and consonants -- e.g., place and manner of articulation.
%\cite{Guenther2009}, 'A potential advantage of decoding articulator positions, rather than formant frequency values, is that consonant production should be easier using a low degree-of-freedom articulatory synthesizer, such as a modified version of the 7-dimensional Maeda articulatory synthesizer [39], than with a formant synthesizer.' / 'In ongoing work we are developing a low-dimensional articulatory synthesizer for use in the BMI.' - 
In the research of Carey et al.~\cite{Carey2017}, MRI of the vocal tract and functional MRI of the brain were recorded with the same speakers, in separate sessions. Since the two signals cannot be recorded simultaneously, the speakers repeated the same stimulus several times for the two modalities, so the relationship between the brain signal and articulation can be examined by aligning through the speech signal.
Chartier and his colleagues~\cite{Chartier2018} inferred the articulatory kinematic trajectories via Acoustic-to-Articulatory Inversion (AAI) methods, from speech acoustics. They used this estimated information to investigate neural mechanisms underlying articulation. The MOCHA-TIMIT database was used to train the AAI model (with speakers independent from the brain signal recordings), in which the articulatory movement was recorded with an electromagnetic articulograph (EMA)~\cite{Wrench2000a}.
Next, in a follow-up~\cite{Anumanchipalli2019}, they estimated the kinematic information about the vocal tract (e.g.~lip movement, tongue movement, and jaw position) as well as other physiological characteristics (e.g.~manner of articulation) from the speech signal, using AAI. They showed that intermediate articulatory representations enhanced performance of speech BCI even with limited data.
Similarly, the aim of~\cite{LeGodais2022} was to analyze the brain signal measured with ECoG device and the articulation information during speech production. He derived the indirect articulatory data from other speakers based on the BY2014 database, using EMA~\cite{Bocquelet2016a}. Since the same sentences but different speakers were used when the brain signal and the articulation signals were recorded, it was possible to calculate EMA-based indirect articulation information for the ECoG data based on the Dynamic Time Warping (DTW) calculated from the speech. Thus, ECoG-based speech synthesis was successfully supplemented with DTW-derived articulation information; although the synthesized speech samples are not yet intelligible \cite{LeGodais2022}.
The latest related study~\cite{Favero2022} also used estimation methods, but here the articulation information is not based on real measurements, but on the so-called TADA features calculated from the speech signal~\cite{Nam2004}.

The conclusion of the above studies is that for patients whose cortical / neural processing of articulation is still intact, a speech-based BCI decoder using articulatory information can be more intuitive or more natural, and easier to learn to use.
%TODO/UltraFest
%\subsection{Goal of the current study}
According to the overview above, there is a lack of combined methods that would examine the non-invasive EEG, articulation, and speech together, and analyze the interaction between the cognitive process in the brain, the articulatory movement, and the speech signal. In the current paper, we extend the analysis of brain signals during speech production with ultrasound tongue image-based articulatory data in order for more biosignals to be available for relationship analysis. From the input brain signal measured with EEG, we use deep neural networks to predict articulatory movement information, in the form of ultrasound tongue image sequences.

\section{Methods}

\subsection{Recordings}

The recordings were made in an electromagnetically shielded quiet room of the ELKH Research Centre for Natural Science, Budapest, Hungary. The EEG signal was recorded with a 64-channel Brain Products actiCHamp type amplifier, using actiCAP active electrodes. Four channels were used to track horizontal and vertical eye movements. The electrodes were placed according to the international 10-20 arrangement~\cite{Klem1999}. The impedance of the electrodes was kept below 15 kOhm. During the recording, the FCz electrode played the role of the reference electrode. The signal was sampled at a frequency of 1000~Hz.

The midsagittal movement of the tongue was recorded using the ,,Micro'' system (AAA v220.02 software, Articulate Instruments Ltd.) with a 2--4~MHz (penetration depth), 64-element, 20~mm radius convex ultrasound probe at 81.67~fps, and we also used a headset for probe fixing. The metal headset was placed above the EEG sensors so that the devices did not interfere with each other. Recording arrangement is shown in Fig.~\ref{fig:felvetel}.

%\begin{figure}
	%\centering
	%\includegraphics[width=0.2\textwidth]{2022-10-06_12-58-09_csapot_EEG.jpg}
	%%\hspace{2mm}
	%\includegraphics[width=0.257\textwidth]{IMG_6878-2_k_Viktor_EEG.jpg}
	%\caption{Recording setup: EEG, ultrasound tongue imaging with a headset, microphone and web camera.}
	%\label{fig:felvetel}
%\end{figure}

The speech was recorded with a Beyerdynamic TG H56c tan omnidirectional condenser microphone and digitized with an M-Audio M-Track 2x2 / FocusRite Scarlett 2i2 USB external sound card at 44,100~Hz. The speaker's face and mouth movements were recorded with a Logitech C925e webcam (but lip data was not used in the current study).

The output of the sound card (which contains the synchronizing signal of the ,,Micro'' ultrasound, i.e., 'frame sync', and the speech signal from the microphone) was connected to the AUX channel of the EEG -- so the brain and articulation signals were recorded on separate computers, but after the session, we can synchronize the data. The EEG signal was recorded continuously, while the ultrasound and speech were recorded sentence-by-sentence. Since the speech signal and the ultrasound synchronization signal (thus the beginning and end of the recording of the given sentence) also appear on one of the EEG channels, we can automatically synchronize the signals afterward. Fig.~\ref{fig:eeg_sync} shows an example of the synchronized signals: EEG (a), speech (b and c), frame sync (d), and ultrasound (e). The latter is a 'kymogram'~\cite[Fig.~8]{Lulich2018}, i.e., a kind of 'articulatory signal over time': the middle slices (midline) of the ultrasound tongue images were cut (approximately corresponding to the middle of the tongue) and plotted as a function of time, similarly to a spectrogram; thus the tongue movement is roughly visible together with the speech spectrogram. %\todo{ábrát nagyobbra tenni}

For the current feasibility study, we recorded approximately 15 minutes of data from a single native Hungarian male speaker (the first author), which will be expanded with additional speakers in the future. The sentences were selected from PPBA~\cite{Csapo2017c}.

\subsection{Preprocessing the data}

\begin{figure}
	\centering
	% trims (crops) from left, bottom, right and top
	\includegraphics[trim=0.0cm 25.8cm 11.5cm 0.0cm, clip=true, width=\columnwidth]{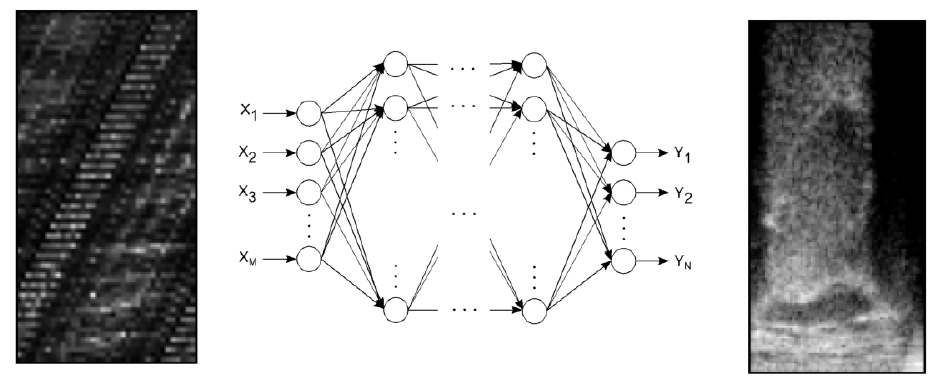}
	%\hspace{2mm}
        \vspace{-4mm}
	\caption{Input (left) and output (right) of the DNNs.}
	\label{fig:EEG_to_UTI}
 \vspace{-4mm}
\end{figure}

The EEG signal was pre-processed based on \cite{Verwoert2022} (\url{https://github.com/neuralinterfacinglab/SingleWordProductionDutch/}). We calculated the Hilbert envelope for each channel of the EEG signal (except EEG AUX) in four frequency bands: 1--50~Hz, 51--100~Hz, 101--150~Hz, and 151--200Hz.  Notch filters were used to filter out the 50~Hz line noise and its harmonics. The envelope was averaged every 50~ms and offset by 12~ms to be consistent with the ultrasound tongue images (which were recorded at 81.67~fps). In order to take temporal information into account, we used 4 preceding and 4 following blocks of the Hilbert-transformed EEG signals. The left side of Fig.~\ref{fig:EEG_to_UTI} shows an example of such an input signal.

Ultrasound tongue images were as 8-bit grayscale pixels, in the form of raw ultrasound of the ,,Micro'' system. The originally 64x842 pixel images were resized to 64x128 pixels (\ref{fig:EEG_to_UTI}.~figure, right side), as this does not cause significant information loss \cite{Csapo2022o}, but the amount of data to be processed is less.

\subsection{Predicting articulatory information from EEG input}

In the current initial phase of the research, we performed a simple experiment: we trained a fully connected (FC-DNN) deep 'rectifier' neural network~\cite{Glorot2011}, during which we predicted the  ultrasound tongue images, from the Hilbert-transformed EEG input (Fig.~\ref{fig:EEG_to_UTI}). For training, MSE error was used. During our experiments, we used a neural network structure with 5 hidden layers, each layer containing 1000 neurons, with ReLU activations, and a linear output layer (similar to earlier ultrasound-based speech synthesis studies, e.g.~\cite{Csapo2017c}). The input EEG values and the output ultrasound pixels were normalized to 0--1 before training. We trained until up to 100 epochs, but applied early stopping, with a patience of 3.

\section{Experiments and results}

We performed training from the 155 sentences, using 80\% of the data for training the network, 10\% for validation, and the remaining 10\% for testing (31\,000 / 3900 / 3900 sample points).

\subsection{Demonstration samples}

\begin{figure}
	\centering
	
	% trims (crops) from left, bottom, right and top
	
    \includegraphics[trim=0.1cm 0.0cm 0.1cm 0.0cm, clip=true, width=\columnwidth]{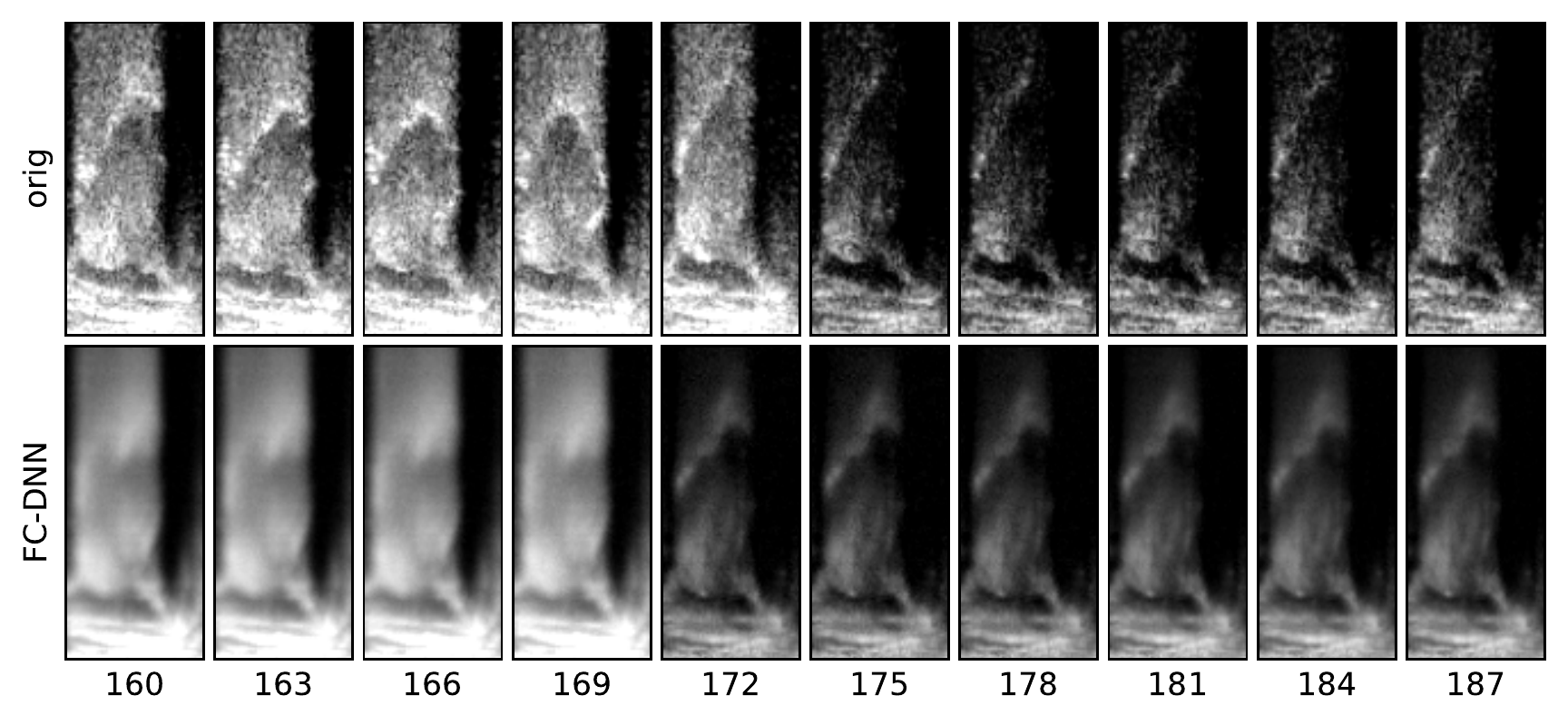}
	%\hspace{2mm}
	\caption{Original (above) and EEG-predicted (below) ultrasound tongue images, in 'raw' representation.}
	\label{fig:EEG_to_UTI_result1}

	% trims (crops) from left, bottom, right and top
        % original
	\includegraphics[trim=0.0cm 0.3cm 4.0cm 0.0cm, clip=true, height=0.73\textheight]{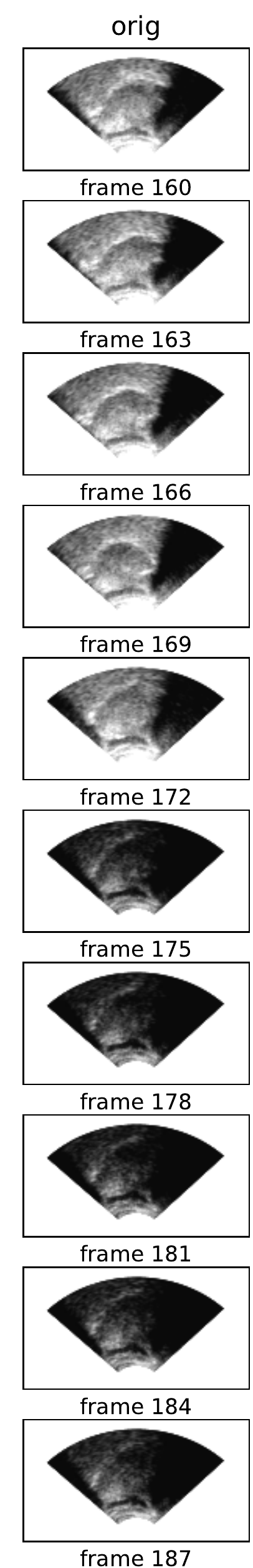}
	\hspace{4mm}
        % FC-DNN
    \includegraphics[trim=4.1cm 0.3cm 0.0cm 0.0cm, clip=true, height=0.73\textheight]{Fig_generated_UTI_sequence_wedge_v2.pdf}
	\caption{Original (left) and EEG-predicted (right) ultrasound tongue images, in 'wedge' representation.}
	\label{fig:EEG_to_UTI_result2}
 
\end{figure}

After the DNN training, ultrasound tongue image prediction was performed from EEG input, on the test set. Fig.~\ref{fig:EEG_to_UTI_result1} shows some original and from EEG estimated ultrasound images from the test data of the speaker, in the 'raw' representation of the ultrasound machine. The contour of the tongue ultrasound is not always visible even in the original images -- this is due to the dependence of the ultrasound tongue images on the speaker -- it seems that this subject has a tongue that is difficult to acquire. In the images estimated from EEG input (i.e., the result of EEG-to-UTI prediction), the contour of the tongue is blurred, and the change in the position of the tongue from frame to frame is also difficult to observe -- i.e., the DNN was able to learn the general shape of the tongue (the average image), but the fine details of the tongue movement cannot be seen. However, some general change of brightness is visible as a function of time: if the original images were darker, then this is also mapped on the predicted images (e.g, around frames 169--172--175). The one image offset in the DNN-predicted image sequence might be the result of windowing the EEG signal.

The same series of images are shown in the 'wedge' representation in Fig.~\ref{fig:EEG_to_UTI_result2}, which were plotted using \url{https://github.com/UltraSuite/ultrasuite-tools}. In these images, a similar trend can be noticed as in Fig.~\ref{fig:EEG_to_UTI_result1}: the upper surface of the tongue can be roughly seen in the original images, but in the images estimated based on the EEG, the ultrasound pixels are blurred, and the contour of the tongue is not visible. However, between frames 169--175, the change in light intensity can be noticed in the DNN-predicated case.

% \begin{figure}
% 	\centering
	
% 	% trims (crops) from left, bottom, right and top
% 	\includegraphics[trim=0.0cm 0.0cm 0.0cm 0.0cm, clip=true, height=0.65\textheight]{Fig_generated_UTI_sequence_wedge_v1.pdf}
% 	%\hspace{2mm}
% 	\caption{Demonstration sample: original (left) and EEG-predicted (right) ultrasound tongue images from speaker FF1, in 'wedge' representation.}
% 	\label{fig:EEG_to_UTI_result2}
% \end{figure}

\begin{figure}
	\centering
	
	% trims (crops) from left, bottom, right and top
	\includegraphics[trim=0.7cm 0.1cm 1.1cm 3.6cm, clip=true, width=\columnwidth]{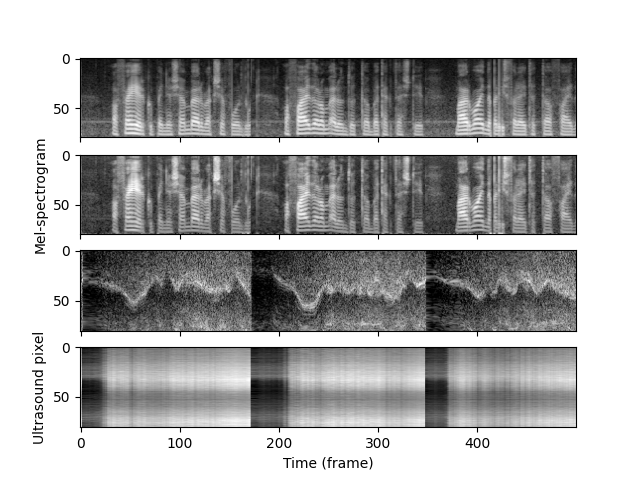}
	%\hspace{2mm}
	\caption{Demonstration sample: a) 80-dimensional mel-spectrogram of the original speech sample, b) original ultrasound kymogram, c) DNN-predicted ultrasound kymogram. \vspace{-5mm}}
	\label{fig:EEG_to_UTI_result3}
\end{figure}

If we look at the results image by image (as in Figs.~\ref{fig:EEG_to_UTI_result1} and~\ref{fig:EEG_to_UTI_result2}), then the longer-term trend is less visible. For this reason, we also show the results in a different arrangement, as a 'kymogram' representation: we cut out the middle vertical line from each ultrasound tongue image and plotted the change of this line over time. Fig.~\ref{fig:EEG_to_UTI_result3} shows the result of this: at the top is the spectrogram belonging to speech, in the middle is the ultrasound image center line sequence as a function of time (belonging to the same utterance), and at the bottom is the ultrasound tongue center line predicted by the DNN. The similarity between a) mel-spectrogram and b) articulatory movement is clearly noticeable: the formant movements in speech and the vertical movement of the tongue can be roughly observed in the figures. On the other hand, in c) DNN-predicted tongue ultrasound, tongue movement is not visible on the midline, i.e., the FC-DNN could not learn well the relation between EEG and ultrasound tongue images. At the same time, some information can still be seen in the DNN-predicated images: at the end of the 170th~frame, one sentence ends, and the next begins, which can be clearly seen in the original ultrasound (b) and also in the estimated ultrasound (c). Overall, we can say that according to this visualization, there is a weak but noticeable relationship between EEG and ultrasound tongue images, i.e. the network can differentiate articulated speech vs.~neutral tongue position.

\subsection{Objective measures}

The mean squared error (MSE) values achieved with the above FC-DNN network are: train error: 0.0052, validation error: 0.0054, test error: 0.0055. The values themselves are difficult to interpret, but for example, in previous UTI-based acoustic-to-articulatory inversion experiments (during which ultrasound tongue images were predicted from the speech signal~\cite{Porras2019,Csapo2021inv}), the obtained NMSE validation error values were in the order of 0.0053--0.0088; and in this case, the ultrasound tongue video generated from the speech input approximated the original articulatory movement. It can also be seen from this that the MSE value in the present research is not sufficient to judge the quality of the results, and a visual inspection is necessary. In the case of previous ultrasound research, there have been experiments examining other error measures, such as Structural Similarity Index (SSIM)~\cite{Wang2004} and Complex Wavelet Structural Similarity (CW-SSIM)~\cite{Sampat2009}, on ultrasound~\cite{Xu2016a,Csapo2020e,Csapo2021inv}. However, due to the above visually weak results in Figs.~\ref{fig:EEG_to_UTI_result1}-\ref{fig:EEG_to_UTI_result2}-\ref{fig:EEG_to_UTI_result3}, we did not investigate SSIM and CW-SSIM here for the case of EEG-to-UTI prediction, as we do not expect they would be helpful.

\section{Discussion and conclusions}

Through the multimodal (brain, speech, and articulation) analysis and synthesis described in this initial research phase, we go beyond the most state-of-the-art international trends.
In our scientific overview of Sec.~1, we have seen many previous attempts for initial speech BCI research based on EEG (or brain signals measured with other invasive devices) ~\cite{Herff2015,Anumanchipalli2019,Krishna2020,Arthur2022a}. However, so far, it has not been possible to create a clearly intelligible synthesized speech  based on brain signals. An obvious solution seems to be the examination of articulation as an intermediate representation between the brain signal and the resulting final speech, which we dealt with in this article. In previous speech-BCI research studies, articulatory information was only included indirectly (i.e., not measured with a piece of equipment), during the investigation of the brain signal and speech~\cite{Guenther2009,Brumberg2011,Carey2017,Chartier2018,Anumanchipalli2019,Favero2022,LeGodais2022}. Although this indirect articulation information also helped to improve the results, measuring and analyzing articulation with real equipment could result in further advantages and improvement in the long-term.

In the current research, we extended the investigation of the brain signal measured with EEG and speech recorded with a microphone, with articulatory recordings using ultrasound tongue imaging. We made sure that all the signals were in good synchrony, by using hardware sync for ultrasound, and by connecting the sync signals and recording the microphone within the EEG equipment as well.
We trained a deep neural network (FC-DNN) to estimate ultrasound images, based on EEG input. According to the results, the generated ultrasound tongue images are still far from the original sequence, but the relationship between EEG and ultrasound tongue images was clearly demonstrated, i.e. the network can differentiate articulated speech  and neutral tongue position, like Voice Activity Detection.

In the future, we plan to compare the prediction results between articulatory and speech data, i.e.~whether the articulatory data can yield better predictions. The long-term goal of this research is to contribute to speech-based brain-computer interfaces. The results can potentially be used during rehabilitation, e.g.~as a communication aid.

The keras implementation of the DNN experiments presented above is available at the following address: \url{https://github.com/BME-SmartLab/EEG-to-UTI}.

\section{Acknowledgements}
\label{sec:ack}

% TODO: later + placeholder. 

% The authors would like to thank ISCA and the organizing committees of past INTERSPEECH conferences for their help and for kindly providing the previous version of this template.

% The authors would like to thank ISCA and the organizing committees of past INTERSPEECH conferences for their help and for kindly providing the previous version of this template.

% \section{Acknowledgements}

This research was funded by the National Research, Development and Innovation Office of Hungary (FK~142163 grant). T.G.Cs.~was supported by the Bolyai János Research Fellowship of the Hungarian Academy of Sciences and by the ÚNKP-22-5-BME-316 New National Excellence Program of the Ministry for Culture and Innovation from the source of the NRDIF. 
% TODO: finalize.

%Szeretnénk köszönetet mondani TODO Lucának és TODO Emesének az EEG felvételekben nyújtott segítségért, az ELKH TTK-nak az EEG eszközök biztosításáért, valamint az MTA-ELTE ,,Lendület'' Lingvális Artikuláció Kutatócsoportnak a nyelvultrahang eszközök rendelkezésre bocsátásáért.

\clearpage

\bibliographystyle{IEEEtran}

\bibliography{ref_collection_csapot_nourl}

\end{document}